\documentclass{article}
\newcommand{\ud}{\mathrm{d}}
\usepackage{amsmath}
\usepackage[dvips]{graphicx}
\begin{document}
\begin{flushright}
MIT-CTP-3519
\end{flushright}
\bigskip
\Large
\begin{center}
{Symmetries of Discontinuous Flows and the Dual Rankine-Hugoniot 
Conditions in Fluid Dynamics}
\end{center}
\bigskip
\bigskip
\normalsize
\begin{center}
{Oliver Jahn\footnote{jahn@mit.edu}}

{Center for Theoretical Physics}

{Massachusetts Institute of Technology}

{Cambridge, MA 02139-4307}

{U.S.A.}
\end{center}
\bigskip
\begin{center}
{V. V. Sreedhar\footnote{sreedhar@iitk.ac.in}}

{Department of Physics}

{Indian Institute of Technology}

{Kanpur, 208016}

{India}
\end{center}
\bigskip
\begin{center}
{Amitabh Virmani\footnote{virmani@physics.ucsb.edu}}

{Department of Physics}

{University of California}

{Santa Barbara, CA 93106-9530}

{U.S.A.}
\end{center}

\bigskip
\bigskip

\begin{abstract}
It has recently been shown that the maximal kinematical invariance group of 
polytropic fluids, for smooth subsonic flows, is the semidirect product of 
$SL(2,R)$ and the static Galilei group $G$. This result purports to offer a 
theoretical explanation for an intriguing similarity, that was recently
observed, between a supernova explosion and a plasma implosion. In this paper 
we extend this result to discuss the symmetries of discontinuous flows, which 
further validates the explanation by taking into account shock waves, which are 
the driving force behind both the explosion and implosion. This is accomplished
by constructing a new set of Rankine-Hugoniot conditions, which follow from 
Noether's conservation laws. The new set is dual to the standard 
Rankine-Hugoniot conditions and is related to them through the $SL(2, R)$ 
transformations. The entropy condition, that the shock needs to satisfy for 
physical reasons, is also seen to remain invariant under the transformations. 
\end{abstract}

\normalsize
\section{Introduction}

It has recently been observed that the density profiles of a supernova 
explosion and an inertial confinement plasma implosion\cite{sn1,sn2,sn3}
are strikingly similar. An empirical basis for this intriguing duality 
between explosion and implosion was given by Drury and Mendon\c{c}a 
\cite{drury} who pointed out that Euler's equations of fluid dynamics, which 
describe both the systems, are form--invariant under a set of nonlinear 
coordinate transformations {\it viz.} $\vec x \rightarrow \vec x/t, ~~~t 
\rightarrow -1/ t$. The minus sign in the time transformation maps an explosion
to an implosion and the inversion allows large time scales to be mapped to 
small time scales and {\it vice versa}. These transformations suggest that the 
maximal kinematical invariance group $\mathcal{G}$ of fluid dynamics is larger 
than the standard Galilei group. It is now known that this larger group is a 
twelve-parameter semidirect product, $\mathcal{G} = SL(2,R)\wedge G$ 
\cite{sreedhar, jackiw}, where $G$ is the nine-parameter, connected, static 
Galilei group:
\begin{equation}
\vec x \rightarrow {\bf R} \vec x + \vec v t + \vec a, \qquad t \rightarrow t
\end{equation}
and $SL(2,R)$ is the group consisting of the transformations:
\begin{equation}
t \rightarrow \frac{\alpha t + \beta}{\gamma t + \delta} \;,
\qquad \vec x \rightarrow \frac{\vec x}{\gamma t + \delta}
\qquad\text{with}\qquad \alpha \delta - \beta \gamma = 1 \;.
\end{equation}
Physically, the three-parameter $SL(2,R)$ group consists of time translations,
scale transformations, and a one-parameter set of time-dependent scale 
transformations called expansions. The transformations proposed by Drury and 
Mendon\c{c}a are a special case of the $SL(2, R)$ transformations with $(\alpha,
~\beta,~\gamma,~\delta) = (0,~-1,~1,~0)$. 
The $SL(2,R)$ part of ${\cal G}$ is therefore important for a better 
understanding of the explosion--implosion map.  

It should be pointed out that the naive expectation of using time-reversal 
invariance, to explain the similarity between explosion and implosion, is 
untenable here since the length and time scales involved in the two systems 
are drastically different. Invoking scaling arguments is not of much help 
since, although a composition of time-reversal and suitable scalings
leaves the equations of fluid mechanics and the Reynolds number invariant, it 
has the property of reversing the direction of time's arrow and thereby 
violates the second law of thermodynamics. As a consequence, when applied to a 
shock wave, such transformations violate entropy conditions that define the 
physicality of the shock. As is well-known, however, both the supernova 
explosion and the plasma implosion are driven by the formation and 
propagation of a shock wave. It is therefore important to examine whether the 
$SL(2, R)$ symmetry, that purports to explain the observed duality, respects 
the physicality of the shock wave. With this in view, we extend the study of 
\cite{sreedhar} -- in which  the explanation of explosion-implosion duality 
based on the symmetry group ${\cal G}$ was restricted to smooth, subsonic 
flows -- to examine shock waves. It will be shown that shock wave solutions 
are consistent with the symmetries of the maximal kinematical invariance group 
${\cal G}$ in the following precise sense.      

A shock in a fluid is described mathematically by the well-known 
Rankine-Hugoniot jump conditions \cite{rich}. So the natural question to ask 
is: What happens to these conditions under the action of the $SL(2, R)$ group? 
This question is best answered not in the framework of the partial differential
equations of fluid dynamics, but by reverting back to their so-called primitive
form {\it i.e.} expressing them as conservation laws. The conservation laws are
completely equivalent to the partial differential equations for smooth flows,
but produce the Rankine-Hugoniot jump conditions for discontinuous flows in a
natural and well-defined manner. The connection with $SL(2, R)$ is made by 
appealing to Noether's theorem which asserts that corresponding to every
continuous symmetry, there exists a conserved charge. 

Anticipating that the $SL(2,R)$ transformations will mix the conservation laws 
corresponding to various symmetries, we construct the Noether charges 
corresponding to them and the boost transformations, in addition to the 
well-known ones for rotations, space and time translations. We then use the 
attendant conservation laws to establish a new set of jump conditions. It turns
out that the new conditions are identically satisfied if the standard 
conditions for mass, momentum and energy conservation are satisfied. Although 
seemingly redundant because of this reason, the new set holds independently; 
following, as it does, from the symmetries of the fluid equations. In fact, 
these conditions are useful to prove the form-invariance of the Rankine-Hugoniot
conditions under the $SL(2,R)$ transformations. Thus, to each physical system 
governed by the fluid dynamics equations two independent, but physically 
equivalent, sets of jump conditions can be associated, the two being related by
$SL(2,R)$ transformations. We conclude that the $SL(2,R)$ transformations map 
the Rankine-Hugoniot conditions of the explosion to the dual Rankine-Hugoniot 
conditions of the implosion and {\it vice versa}. Further, by specialising 
to the Drury-Mendon\c{c}a transformations, $\vec x \rightarrow \vec x/t, ~~~t
\rightarrow -1/ t$, we show that the jump conditions for boosts and 
expansions, along with the continuity equation for mass conservation, provide
an independent, albeit equivalent, description of the shock. They may 
be viewed either as the dual of the standard Rankine-Hugoniot conditions, or, 
in the language of passive coordinate transformations, as the standard 
Rankine-Hugoniot conditions in the dual coordinate system corresponding to the 
choice $(\alpha, \beta, \gamma, \delta ) = (0, -1, 1, 0)$. Similar dual 
conditions exist for each choice of the $SL(2, R)$ parameters. 

It is well-known that Rankine-Hugoniot conditions describe not only shocks, but
other discontinuities like slip and contact discontinuities, for example. 
Therefore, the map between the dual sets of Rankine-Hugoniot conditions would 
be relevant to explosion-implosion duality only if both the sets refer to 
shocks. In other words, only those Rankine-Hugoniot conditions that describe a 
shock and only those $SL(2, R)$ transformations which map a shock to a shock 
are of interest for explosion-implosion duality. Moreover, Rankine-Hugoniot 
conditions say nothing about the physicality of the shock -- this 
information is contained in additional inequalities for its entropy that the 
shock needs to satisfy. Physical shocks are distinguished from others because 
their entropy always increases across the shock front. We verify explicitly 
that this requirement is unaffected by the $SL(2, R)$ transformations. 

The physicality of the map between explosion and implosion may also be 
established in the following subtle manner: Although the notions of viscosity 
and heat conduction lose their meaning in the immediate vicinity of the shock, 
because the changes in all the quantities they depend on are so great, they do 
play an important role in the formation and maintenance of a shock 
discontinuity\cite{courant}. In particular, the positivity of the coefficients 
of viscosity and heat conduction guarantees that the shock satisfies the 
appropriate entropy conditions\cite{courant, landau}. Hence, Euler's equations 
ought to be considered as a special case of a more general set of fluid 
equations with vanishingly small viscosity. Requiring the sign of the viscosity
to remain unchanged under the transformations establishes the physicality. The 
Navier-Stokes equations -- which are the obvious choice for 
including viscosity --  are not invariant under the full $SL(2,R)$ part
of ${\cal G}$, but only under the standard Galilean transformations. However,
a more general set of fluid equations with viscosity fields transforming
appropriately under the $SL(2, R)$ transformations has a maximal kinematical
invariance group given by ${\cal G}$ \cite{sreedhar}. Hence we use these 
equations for our purpose of examining the behaviour of non--vanishing 
viscosity under the $SL(2, R)$ transformations. Similar 
arguments apply for heat conduction, but it does not bring in any new 
qualitative features and hence is omitted from further discussion. 

\section{Symmetries of Fluid Dynamics}

In this section we briefly recapitulate the results of \cite{sreedhar}.
The general fluid equations in $n$-dimensional space are \cite{landau}
\begin{eqnarray} \label{e1}
D \rho &=& - \rho \vec{\nabla} \cdot \vec{u} 
\\\label{e2}
\rho D \vec{u} &=&  -\vec{\nabla} p + \vec{V} 
\\\label{e3}
D \varepsilon &=&  - (\varepsilon + p) \vec{\nabla} \cdot \vec{u} 
\end{eqnarray}
where
\begin{displaymath}
D = \frac{\partial}{\partial t} + \vec{u} \cdot \vec{\nabla}
\end{displaymath}
and
\begin{displaymath}
V_i = \nabla_j\left(\eta(\nabla_j u_i + \nabla_i u_j - \frac{2}{n} \delta_{ij} \nabla_k u_k)\right) + \nabla_i(\zeta \nabla_k u_k)
\end{displaymath}
In the above equations $\rho, \vec{u},p,\varepsilon$ stand for the
density, velocity, pressure and energy density of the
fluid respectively and $\eta, \zeta$ are the bulk and shear viscosity fields.

The above differential equations are usually augmented by an algebraic 
condition called the polytropic equation of state which relates the pressure 
to the energy density as  
\begin{equation}\label{eos}
p = (\gamma_0 - 1) \varepsilon
\end{equation}
where $\gamma_0$ is a constant called the polytropic exponent. As shown in 
\cite{sreedhar}, the maximal invariance group of the above set of equations is 
${\cal G} = SL(2, R)\wedge G$, provided the polytropic exponent takes the 
standard value for an ideal, nonrelativistic fluid {\it viz.} $\gamma_0 = 1 + 
\frac{2}{n}$. For this value, the fluid equations are invariant under the 
following transformations \cite{sreedhar}:

\subsubsection*{Connected, static Galilei transformations:}
Let $g$ denote a general element of this sub-group then 
\begin{equation}
g: \quad t' = t \;,\qquad \vec x' = {\bf R} \vec x + \vec v t + \vec a 
\end{equation}
with ${\bf R}$ an orthogonal matrix.  Under the action of $g$, the fields 
$\rho$ and $\vec u$ transform as
\begin{equation}\label{field1}
\rho' = \rho \quad\mbox{and}\quad  
\quad 
\vec u ' = \vec u  + \vec v
\end{equation} 
\subsubsection*{SL(2,R) transformations:}
Let $\sigma$ denote a general element of the $SL(2,R)$ part of $\mathcal{G}$ 
then
\begin{equation}\label{cor2}
\sigma : \quad t' = \frac{\alpha t + \beta}{\gamma t + \delta} \;,\quad
\vec x' = \frac{\vec x}{\gamma t + \delta}
\quad\text{where}\quad \alpha \delta - \beta \gamma  = 1
\end{equation}
Under the action of $\sigma$, the fields transform as
\begin{eqnarray}\label{field2}
\rho' = (\gamma t + \delta)^n  \rho\quad\mbox{and}\quad 
\vec u ' = (\gamma t + \delta)\vec u - \gamma\vec x  
\end{eqnarray}
For both $g$ and $\sigma$, the transformations of $\varepsilon$ and $p$ can 
be worked out once the transformation properties of $\rho$ are known since
\begin{equation}
  \label{rel}
  \varepsilon= \chi\rho^{\gamma_0}\;, \qquad
  p=(\gamma_0-1)\varepsilon\;, 
\end{equation}
with the field $\chi$ -- related to entropy -- transforming like a scalar. 
The transformation properties of the viscosity fields are similar to 
the density $\rho$ 
\begin{equation}
  \eta' = \bigl(\gamma t+ \delta\bigr)^n\eta \quad\mbox{and}\quad  
  \zeta' = \bigl(\gamma t+ \delta\bigr)^n\zeta  
\end{equation} 
The above results were derived in \cite{sreedhar} by requiring the invariance
of the Action for the simple case of an inviscid and isentropic fluid. The 
symmetry of the equations followed by subsequently relaxing the simplifications
to arrive at the general fluid equations. It should be noted that the 
requirement of the invariance of the Action is sufficient, but not necessary, 
for the form invariance of the equations that follow from it. Any 
transformation that leaves the Action invariant upto a multiplicative factor
produces equations of motion which have the same form. If this 
is taken into account, the condition $\alpha\delta - \beta\gamma 
= 1$ is no longer required and $SL(2, R)$ gets replaced by $GL(2, R)$ in the 
maximal invariance group of the general fluid equations. However, it is 
sufficient for our purposes to concentrate on the variational symmetries of 
the fluid equations and for this purpose, $\mathcal{G} = SL(2,R)\wedge G$. 
Finally, for the sake of completeness, it should also be pointed out that the 
$SL(2,R)$ condition is invariant under the following discrete symmetries
$(\alpha,\beta, \gamma,\delta)~~\rightarrow~~ (\alpha,-\beta,-\gamma,\delta), 
~~\rightarrow~~ (-\alpha,\beta,\gamma,-\delta), ~~\rightarrow~~ (-\alpha,
-\beta,-\gamma, -\delta)$ 
of the $SL(2,R)$ parameters. 

\section{Conservation Laws}

In this section we construct the conservation laws corresponding to the 
symmetries outlined in the previous section. In order to do this, it is 
useful to revert back to the Action formalism and obtain the results for the 
subclass of inviscid, isentropic and irrotational flows. The corresponding 
expressions for a general fluid can then be worked out along the lines of 
~\cite{sreedhar}. 

For inviscid, isentropic and irrotational flows, the Lagrangian density is 
given by 
\begin{equation}
{\cal L} = \rho\bigl(\dot\phi - {1\over 2}(\vec\nabla\phi)^2\bigr) - 
\rho^{\gamma_0} 
\end{equation} 
where $\vec\nabla\phi$ stands for the curl-free part of the velocity vector
field $\vec u$. Let $\mu = 0,1,2,3$ and $x^\mu$ be a four-vector under the 
transformations of the previous section {\it i.e.} $x^i$ with $i = 1,2,3$ are 
the components of $\vec x$  and  $x^0 = t$. Let the infinitesimal variations in
the coordinates and fields be defined as   
\begin{equation}
\delta x^\mu = {x^\mu}' - x^\mu \quad\mbox{and}\quad \delta\phi (x) = \phi'(x') 
- \phi (x)
\end{equation} 
Then the variations for translations, rotations, boosts, dilatations, and 
expansions respectively are given by 
\begin{equation}
\delta x^\mu = a^\mu,~~ \delta x^i = \omega^{ij}x^j,~~ \delta x^i = v^it,~~ 
\delta x^i = \lambda x^i,~~ \delta t = 2\lambda t~~\mbox{and}~~ 
\delta x^\mu = -\mu tx^\mu
\end{equation}
where the parameters $\lambda,~a^0,~\mu$ are expressible in terms of the 
$SL(2,R)$ parameters $\alpha,~\beta,~\gamma,~\delta$. The field variation 
is given by 
\begin{equation}
\delta\phi = \Lambda = {[\gamma(x + a) -\delta v]^2\over 2\gamma(\gamma t 
+\delta )}
\end{equation}
The variation in $\rho$ is not important since no derivatives of $\rho$ 
appear in the Lagrangian density. Using these results we find, by a 
straightforward application of Noether's theorem \cite{goldstein}, that 
the following quantities, integrated over all space, are constants of motion:
\begin{equation}
{\hbox{\bf Temporal Translations}}:\qquad 
H = {\rho\over 2}(\vec\nabla\phi )^2 + \rho^{\gamma_0}
\end{equation}
\begin{equation}
{\hbox{\bf Spatial Translations}}:\qquad 
\vec P  = \rho\vec\nabla\phi 
\end{equation}
\begin{equation}
{\hbox{\bf Rotations}}:\qquad 
\vec L  = \vec P \times \vec x 
\end{equation}
\begin{equation}
{\hbox{\bf Boosts}}:\qquad 
\vec K  = \vec P t  - \rho\vec x  
\end{equation}
\begin{equation}
{\hbox{\bf Dilatations}}:\qquad 
D  = -2t H + \vec x \cdot \vec P 
\end{equation}
\begin{equation}
{\hbox{\bf Expansions}}:\qquad 
A  = t^2 H - t \vec x\cdot\vec P 
+ {\rho\over 2}{\vec x}^2
\end{equation}
The conditions of irrotationality and isentropicity can be relaxed easily 
and one sees that Euler's equations
\begin{equation}
\dot\rho = -\vec\nabla\cdot(\rho\vec u)
\end{equation}
\begin{equation}
\rho {\dot{\vec u}} = -\rho(\vec u\cdot\vec\nabla)\vec u - \vec \nabla p 
\end{equation}
\begin{equation}
\dot\varepsilon = -\vec\nabla\cdot(\varepsilon\vec u) - p \vec\nabla 
\cdot\vec u 
\end{equation}
can be expressed in the form of conservation laws, 
\begin{equation} 
{\partial\over\partial t}\rho = - {\partial\over\partial x_j} (\rho u_j) 
\end{equation} 
\begin{equation} 
{\partial\over\partial t}(\rho u_i) = - {\partial\over\partial x_j} (\rho 
u_iu_j +\delta_{ij}p) 
\end{equation} 
\begin{equation} 
{\partial\over\partial t}({1\over2}\rho\vec u^2 +\varepsilon ) = - 
{\partial\over\partial x_j} [({1\over 2}\rho\vec u^2 + \varepsilon + p)u_j] 
\end{equation} 
for mass and the translation generators found above. These can be reexpressed 
succinctly as follows:
\begin{equation}
\partial_\mu J^\mu_{(\rho)} = 0,\quad \partial_\mu J^\mu_{(\vec P)} = 0, 
\quad\mbox{and}\quad \partial_\mu J^\mu_{(H)} = 0 
\end{equation}
The zeroeth components of the above currents, namely $\rho$, $\rho\vec u$,
and ${1\over 2}\rho\vec u^2 +\varepsilon$, give the charge densities which, 
when integrated over all space, give the conserved charges. As is well-known, 
these are merely statements of mass, momentum flux, and total energy 
conservation. The corresponding current densities are 
\begin{equation} 
J^j_\rho = \rho u_j
\end{equation}
\begin{equation} 
J^j_{P_i} = \rho u_iu_j + \delta_{ij}p
\end{equation}
\begin{equation} 
J^j_{H} = ({1\over 2}\rho\vec u^2 + \varepsilon + p)u_j
\end{equation}
The conservation laws corresponding to rotations, boosts, dilatations and 
expansions can be stated similarly
\begin{equation}
\partial_\mu J^\mu_{(\vec L)} = 0,\quad \partial_\mu J^\mu_{(\vec K)} = 0, 
\quad \partial_\mu J^\mu_{({D})}, 
 \quad\mbox{and}\quad \partial_\mu J^\mu_{({A})} = 0 
\end{equation}
The charge densities are shown in (19) -(22) respectively, and the 
corresponding currents are 
\begin{equation} 
\vec J_{L_i} = \epsilon_{ikl}x_k\vec J_{P_l}
\end{equation}
\begin{equation} 
\vec J_{K_i} = t\vec J_{P_i} - x_i\vec J_\rho
\end{equation}
\begin{equation} 
\vec J_{D} = x_i\vec J_{P_i} - 2t\vec J_{H} 
\end{equation}
\begin{equation} 
\vec J_{A} = {1\over 2}\vec x^2\vec J_\rho - tx_i\vec J_{P_i} + 
t^2\vec J_{H} 
\end{equation}
It may be mentioned that the above results are not surprising in the light of 
\cite{ajp}, where corresponding results for a free, nonrelativistic, point 
particle were obtained through a discussion that essentially parallels the 
above. The noteworthy linear relations between the currents will, however, play
a crucial role in this paper when we consider flows with discontinuities. 

\section{Discontinuous Flows and Jump Conditions}

As long as the flows are smooth, {\it i.e.} the functions $\rho, \vec{u}, p,
\varepsilon \in \mathbf{C}^1$ in their dependence on $\vec x$ and $t$, the
systems (23 -- 25) and (26 -- 28) are equivalent. However, real flows are not 
always smooth and can develop discontinuities as time elapses. Such flows 
are described by weak solutions of differential equations \cite{rich}. 
A weak solution is generally piecewise smooth. The smooth parts satisfy
the differential equation in the usual, or strong, form, but that does not
generally suffice to determine the course of motion for initial data,
and the equation must be supplemented by jump conditions. The resulting jump 
conditions are most clearly derived from the conservation laws. 

By definition any, possibly non-smooth, function $J^\mu(\vec{x},t)$ that 
satisfies
\begin{equation}\label{weak} \int \partial_\mu w(\vec x,t) \, J^\mu(\vec x,t)
\,\ud^3x\, \ud t = 0 \end{equation}
for all test functions $w(\vec{x},t)$ is said to be a weak solution of
the differential equation $\partial_\mu J^\mu=0$.

We now use the above definition to obtain the jump conditions associated with 
the system of conservation laws derived in the last section. Suppose
$J^\mu(\vec x,t)$ has a jump discontinuity across a hyper-surface
$\mathcal{S}$ in ${\vec x},t$ space, while otherwise being
continuously differentiable in some neighbourhood $\mathcal{N}$ of
$\mathcal{S}$ (see Fig.1).  
\begin{figure}[h]
\begin{center}
\includegraphics[width=0.5 \textwidth]{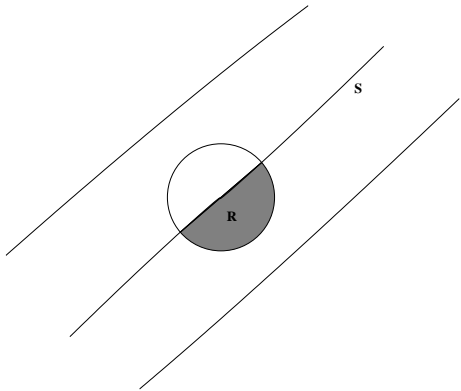}
\end{center}
\caption{Diagram for the jump condition.}
\end{figure}
Let $w(\vec x,t)$ be a test function with support in $\mathcal{N}$. Let 
$\mathcal{R}$ be the part of the support of $w(\vec x,t)$ that lies on 
\emph{one side} of $\mathcal{S}$, say the right. Then, by Gauss's theorem 
\begin{equation}
  \label{tough}
  \int_{\mathcal{R}} \partial_\mu w \,J^\mu \,\ud^3 x\,\ud t 
  + \int_{\mathcal{R}} w \,\partial_\mu J^\mu \,\ud^3 x\,\ud t 
  = \int_{\mathcal{R}} \partial_\mu (w J^\mu) \,\ud^3 x\,\ud t 
  = \int_{\mathcal{S}} w n_\mu J^\mu \,\ud^3 \mathcal{S}
\end{equation}
since $w(\vec x,t) = 0$ on the boundary of $\mathcal{R}$ except on 
$\mathcal{S}$. The second integral in the above equation is zero, because the 
conservation law holds in the strong sense in the interior of $\mathcal{R}$.  
Here, $n(\vec x,t)$ is the outward normal vector to the hypersurface 
$\mathcal{S}$. Therefore, if we integrate similarly over the left part of 
the support of $w$, add the result and make use of (\ref{weak}), we find that:
\begin{equation}
0 = \int_{\mathcal{S}} w n_\mu \Delta J^\mu \ud \mathcal{S}  
\end{equation}
where $\Delta f$ denotes the difference of the two limiting values of a
function $f$ on the two sides of the hypersurface $\mathcal{S}$ {\it i.e.} 
the jump of the function. This result follows because the vector $n_\mu$, which 
by convention points outwards, flips its sign on the left side of the support. 
Since $w$ is an arbitrary test function, the above equation implies the jump 
condition
\begin{equation}\label{master}
n_\mu \Delta J^\mu = 0 \quad\mbox{on } \mathcal{S} \;.
\end{equation}
Applying (\ref{master}) to the conservation laws (29) for $J^\mu_{(\rho)}$, 
$J^\mu_{(\vec P)}$ and $J^\mu_{(H)}$, we obtain
\begin{align}
  0 &= n_\mu \Delta J_{(\rho)}^\mu \;, \label{RH1} \\
  0 &= n_\mu \Delta J_{(\vec P)}^\mu  \;, \label{RH2}\\
  0 &= n_\mu \Delta J_{(H)}^\mu \label{RH3}
\end{align}
From here the standard Rankine-Hugoniot conditions can be derived in their 
usual form \cite{rich}. Similarly, one can apply (\ref{master}) to the 
conservation laws (33) for $J^\mu_{(\vec L)}$, $J^\mu_{(\vec K)}$, 
$J^\mu_{(D)}$, and $J^\mu_{(A)}$ to obtain a new set of jump conditions:
\begin{align}
  0 &= n_\mu \Delta J^\mu_{(\vec L)} \\
  0 &= n_\mu \Delta J_{(\vec K)}^\mu \\
  0 &= n_\mu \Delta J_{(D)}^\mu \\
  0 &= n_\mu \Delta J_{(A)}^\mu .
\end{align}
Since the coordinates $\vec x$ and $t$ are continuous on $\mathcal{S}$, these 
conditions are all identically satisfied because of the jump conditions 
for mass, momentum and energy conservation, in (42-44) -- a fact that can be 
easily verified using (19-22) and (34-37). 

\section{The Dual Rankine-Hugoniot Conditions}

We have seen that the new set of jump conditions associated with rotations, 
boosts, dilatations, and expansions, follow from the jump conditions 
associated with mass, momentum and energy. This suggests that the 
Rankine-Hugoniot conditions are invariant under the full kinematical invariance
group of smooth flows, including the $SL(2,R)$ part.

To see this explicitly, we consider the transformation properties of 
the conserved currents under $SL(2,R)$. Let us begin by considering the 
simplest of these, namely the time-component of $J^\mu_{(\rho )},~i.e.~\rho$.  
From equation (10), now with $n = 3$, 
\begin{eqnarray}\label{density}
\rho' = (\gamma t + \delta)^3  \rho  
\end{eqnarray}
The $(\gamma t + \delta)^3$ factor is cancelled by an identical factor coming
from the change of variables when we perform an integration over all space. 
Moreover, the transformation does not mix $\rho$ with any other current. Thus,  
$\rho$ transforms under the singlet representation of $SL(2, R)$ as a scalar 
density. Let us now consider the transformation of the time-component of 
$J^\mu_{({\vec P})},~i.e.~\vec P = \rho\vec u$. From (10) it now follows,
after a little algebra, that
\begin{eqnarray}\label{momentum}
\vec P' = \rho'\vec u' = (\gamma t + \delta)^3(\delta \vec P + \gamma\vec K)
\end{eqnarray}
Thus the transformation of the spatial translation generator mixes it with the 
boost generator together with which, it forms a doublet representation of 
$SL(2, R)$, with the prefactor $(\gamma t + \delta)^3$ now making it a vector 
density. The latter fact is, in fact,  generic to the time-components of all 
the currents. Likewise, we may consider the generator of time translations, 
namely the Hamiltonian, and it follows that
\begin{eqnarray}\label{Hamiltonian}
\ H' =  (\gamma t + \delta)^3(\gamma^2 A - \delta\gamma D + \delta^2 H)
\end{eqnarray}
Thus the transformation of the time translation generator mixes it with the 
generator of dilatations and expansions, the three of them form the triplet 
(or adjoint) representation. The transformation properties of the rest of the 
currents can be similarly worked out and the results summarised as follows: If 
the (abstract) symmetry generators $T_r$ transform as
\begin{equation}
  T'_r \equiv \sigma^{-1} T_r \sigma =
  \sum_s M_{r s}(\sigma) T_s \;,
\end{equation}
where the matrix  $M(\sigma)$ is determined by the group structure of
$SL(2,R)\wedge G$, then the corresponding currents transform as
\begin{equation}
  J^{\mu\prime}_r(x') = \det\left(\frac{\partial x}{\partial x'}\right)
  \frac{\partial x^{\mu\prime}}{\partial x^\nu}
  \sum_s M_{r s}(\sigma) J^\nu_s(x) \;.
\end{equation}
Assembling the currents in a column, 
\begin{equation}
  J^\mu = \left(\begin{matrix} J^\mu_{(\rho )} \\ J^\mu_{(\vec K)} \\ 
J^\mu_{(\vec P)} \\ J^\mu_{(A)} \\ J^\mu_{(D)} \\ J^\mu_{(H)}
      \end{matrix}\right)
\end{equation}
one has for the transformation matrix, 
\begin{equation}
  M = \left(\begin{matrix}
  1 & 0 & 0 & 0 & 0 & 0 \\
  0 & \alpha & \beta & 0 & 0 & 0 \\
  0 & \gamma & \delta & 0 & 0 & 0 \\
  0 & 0 & 0 & \alpha    & -\alpha\beta      & \beta^2 \\
  0 & 0 & 0 & -2\alpha\gamma & (\beta\gamma + \alpha\delta) & -2\beta\delta \\
  0 & 0 & 0 & \gamma^2   & -\gamma\delta  & \delta^2
  \end{matrix}\right)
\end{equation}
Using $\alpha\delta - \beta\gamma = 1$, and the fact that the determinant of 
a block diagonal matrix is the product of the determinants of the blocks, it is
easy to check that the matrix $M$ has unit determinant. As already pointed out, 
the fact that the currents transform like vector densities is reflected in 
the temporal components picking up a multiplicative factor 
$(\gamma t + \delta)^3$.  The spatial components follow the example 
\begin{equation}
J_{(\rho)}^{i\prime} = (\gamma t+\delta)^{n+1} J_{(\rho)}^i
  - \gamma x^i (\gamma t+\delta)^n J_{(\rho)}^0
\end{equation}
with the same $SL(2, R)$ transformations defined by the matrix $M$. 

The dual Rankine-Hugoniot conditions are now easily obtained. The normal 
vector $n_\mu$ appearing in the jump condition (\ref{master}) transforms like 
a covector,
\begin{equation}
  n_\mu' \propto \frac{\partial x^{\nu}}{\partial x^{\mu\prime}} n_\nu \;,
\end{equation}
so the transformed jump condition for $J_r$ is
\begin{equation}
  n_\mu' \Delta J^{\mu\prime}_r \propto
  \det\left(\frac{\partial x}{\partial x'}\right)
  \sum_s M_{r s}(\sigma) n_\mu \Delta J^\mu_s(x)
  = 0 \quad\mbox{on } \mathcal{S} \;.
\end{equation}
Since the determinant is smooth across the surface $\mathcal{S}$, the
factor in front of the sum can be omitted.  The transformed jump
condition is therefore a linear combination of the original jump
conditions.  In particular, the conditions for $J_{(\rho)}$, $J_{(\vec P)}$ 
and $J_{(H)}$ (the Rankine-Hugoniot conditions) become linear combinations of 
the jump conditions for $J_{(\rho)}$, $J_{(\vec P)}$, $J_{(\vec K)}$, 
$J_{(H)}$, $J_{(D)}$ and $J_{(A)}$,
\begin{align}
n_\mu' \Delta J^{\mu\prime}_{(\rho)} &\propto n_\mu \Delta J^\mu_{(\rho)} \;, \\
  n_\mu' \Delta J^{\mu\prime}_{(\vec P)}
  &\propto n_\mu ( \gamma \Delta J^\mu_{(\vec K)}
+ \delta \Delta J^\mu_{(\vec P)} ) \;, \\
  n_\mu' \Delta J^{\mu\prime}_{(H)}
 &\propto n_\mu ( \delta^2 \Delta J^\mu_{(H)} - \gamma\delta \Delta J^\mu_{(D)}
                 + \gamma^2 \Delta J^\mu_{(A)} ) \;.
\end{align}
The standard Rankine-Hugoniot conditions (42-44), in conjunction with the 
new set of jump conditions (45-48), then imply that the right hand side of 
the above equations is identically zero {\it i.e.} the Rankine-Hugoniot 
conditions are form--invariant. In particular, this holds for the 
Drury-Mendon\c{c}a transformation $t\to-1/t$, $\vec x\to\vec x/t$ used to 
relate the explosion and implosion problems. For this, $(\alpha,\beta, \gamma, 
\delta)  = (0, -1, 1, 0)$ and it follows that  
\begin{align}
  0 &= n_\mu \Delta J^\mu_{(\rho)} \;, \label{dualRH1} \\
  0 &= n_\mu \Delta ( x_i J_{(\rho)}^\mu - t J_{(P_i)}^\mu ) \;, \label{dualRH2}\\
  0 &= n_\mu \Delta ( -t^2 J_{(H)}^\mu + t x_i J_{(P_i)}^\mu - \tfrac12 x^2 
\label{dualRH3} J_{(\rho )}^\mu ) \;.
\end{align}
where we have substituted the explicit expressions for the currents $J^\mu_{
(\vec K)}$ and $J^\mu_{(A)}$. The conditions (62-24) are the dual 
Rankine-Hugoniot conditions. If an explosion is described by the standard 
Rankine-Hugoniot conditions, the corresponding implosion, obtained by a 
Drury-Mendon\c{c}a transformation, is described by the dual 
Rankine-Hugoniot conditions (62-64).

Since the coordinates $\vec x$ and $t$ are continuous on $\mathcal{S}$, and 
crucially because the relations between the currents are linear, the conditions
(62-64) are equivalent to the jump conditions obtained from mass, momentum and 
energy conservation, in (42-44). In fact, these two sets of equations imply, 
and are implied by, each other. In conclusion, the dual set of jump conditions 
associated with mass, boosts and expansions, is completely equivalent to the 
usual Rankine-Hugoniot conditions and may be used for an independent 
description of the shock. 

\section{The Entropy Condition}
 
For a polytropic gas, by choosing $\varepsilon = \chi \rho^{\gamma_0}$, we can 
rewrite Eq. (5) as 
\begin{equation}
D\chi = 0
\end{equation}
In \cite{sreedhar} we defined an isentropic flow to be one for which 
$\chi =~constant$. For a general flow, it followed that $\chi$ transforms like 
a scalar. For a polytropic gas, it is also well-known \cite{courant} that 
$\chi$ is related to the specific entropy (entropy per unit mass), $S$ as 
follows:
\begin{equation}
S-S_0  = C_v{\hbox{log}}\bigl[\chi (\rho V)^{\gamma_0}\bigr] 
\end{equation}
where $C_v = R/(\gamma_0 - 1)$, $R$ being the universal gas constant 
divided by the molecular weight of the particular gas, $V$ the volume and 
$S_0$ an appropriate constant. It is obvious from this equation that as a 
particle of the medium moves about, the specific entropy at the moving particle
remains constant under an $SL(2, R)$ transformation. Hence, under an $SL(2, R)$ 
transformation, a physical shock gets mapped to a physical shock. 
 
We now require the positivity of viscosity to be preserved under an 
$SL(2, R)$ transformation -- a requirement that guarantees that the shock 
respects the entropy condition. As already pointed out (see eq. (12)), 
in three-dimensional space, the viscosity fields transform as follows: 
\begin{equation}
  \eta' = \bigl(\gamma t+ \delta\bigr)^3\eta \quad\mbox{and}\quad
  \zeta' = \bigl(\gamma t+ \delta\bigr)^3\zeta
\end{equation}
Thus the transformation properties of the viscosity fields are similar to 
$\rho$ {\it i.e.,} they transform like scalar densities. Hence, if we 
integrate the viscosity field over all space, to get the viscosity, it is 
an invariant under the $SL(2, R)$ transformations. Likewise, the specific 
viscosity (viscosity per unit mass), is an invariant. It follows that the 
positivity of the viscosity is maintained without any additional restrictions 
on the  $SL(2, R)$ parameters.

\section{Conclusions}

In this paper, we extended the analysis of \cite{sreedhar} to discuss the 
symmetries of discontinuous flows in fluid dynamics. The maximal kinematical 
invariance group of an ideal, polytropic fluid is ${\cal G}=SL(2,R)\wedge G$, 
not just for smooth, but for discontinuous flows also. This is made manifest 
by writing the fluid equations in their conservation law form. New conservation
laws follow from a direct application of Noether's theorem, enabling us to 
construct a dual set of Rankine-Hugoniot shock conditions. The $SL(2, R)$ 
transformations map the standard Rankine-Hugoniot shock conditions to the dual 
ones and {\it vice versa}. These transformations also respect the entropy
conditions that physical shocks need to satisfy. Hence we conclude that, under 
these transformations, an explosion gets mapped to an implosion, thus offering 
a theoretical explanation for the intriguing observations of \cite{sn1,sn2,sn3}.

\section*{Acknowledgements} We thank L. O'C Drury, Pravir Dutt and S.G. Rajeev
for discussions. VVS thanks A. J. Niemi for his hospitality in Uppsala, Sweden,
and the Center for Dynamical Processes and Structure Formation (CDP),  
Uppsala University, for financial support.

\end{document}